\documentclass[aps,prd,letterpaper,twocolumn,preprintnumbers,floatfix,superscriptaddress]{revtex4}

\pdfoutput = 1

\usepackage[usenames,dvipsnames]{color}
\usepackage{graphicx}
\usepackage{amsmath,amsfonts,amssymb,dsfont,bm,bbm,dcolumn,float,ifthen,mathrsfs,wasysym}
\usepackage{pstricks,feyn,verbatim}
\usepackage[colorlinks=true,linkcolor=blue,citecolor=blue,urlcolor=black]{hyperref}
\usepackage{natbib}

\usepackage{slashed,xcolor,multirow,array,rotating}
\usepackage{fancybox,epstopdf}

\newcommand{\GeV}      {~\mathrm{GeV}}
\newcommand{\TeV}      {~\mathrm{TeV}}

\newcommand{\fb}      {~\mathrm{fb}}

\newcommand{\Tr}   {~\mathrm{Tr}}

\newcommand{\ba}{\begin{array}}
\newcommand{\ea}{\end{array}}
\newcommand{\beqn}{\begin{eqnarray}}
\newcommand{\eeqn}{\end{eqnarray}}
\newcommand{\beqs}{\begin{subequations}}
\newcommand{\eeqs}{\end{subequations}}
\newcommand{\be}{\begin{equation}}
\newcommand{\ee}{\end{equation}}

\newcommand{\mathsym}[1]{{}}

\def\gU{\rm U}
\def\gSU{\rm SU}

\def\gSp{\rm Sp}

\def\mL{\mathcal{L}}

\def\mO{\mathcal{O}}

\def\hf{\frac{1}{2}}

\usepackage{ulem,fancyvrb}

\begin{document}
\title{A hidden confining world on the 750 GeV diphoton excess}

\author{Ligong Bian}
\affiliation{Department of Physics, Chongqing University, Chongqing 401331, China}
\affiliation{Kavli Institute for Theoretical Physics China, Institute of Theoretical Physics, Chinese Academy of Sciences, Beijing 100190, P. R. China}
\author{Ning Chen}
\email{chenning@ustc.edu.cn}
\affiliation{Department of Modern Physics, University of Science and Technology of China, Hefei, Anhui, 230026, China}
\author{Da Liu}
\affiliation{Kavli Institute for Theoretical Physics China, Institute of Theoretical Physics, Chinese Academy of Sciences, Beijing 100190, P. R. China}
\author{Jing Shu}
\email{jshu@itp.ac.cn}
\affiliation{Kavli Institute for Theoretical Physics China, Institute of Theoretical Physics, Chinese Academy of Sciences, Beijing 100190, P. R. China}
\affiliation{CAS Center for Excellence in Particle Physics, Beijing 100049, China}

\begin{abstract}

We explain the recent diphoton excesses around $750$ GeV by both ATLAS and CMS as a singlet scalar $\Phi$ which couples to SM gluon and neutral gauge bosons only through higher dimensional operators. 
A natural explanation is that $\Phi$ is a pseudo-Nambu-Goldstone boson (pNGB) which receives parity violation through anomaly if there exists a hidden strong dynamics. 
The singlet and other light pNGBs will decay into two SM gauge bosons and even serves as the meta-stable coloured states which can be probed in the future. 
By accurately measuring their relative decay and the total production rate in the future, we will learn the underlying strong dynamics parameter. 
The lightest baryon in this confining theory could serve as a viable dark matter candidate.

\end{abstract}


\maketitle



\section{Introduction}
\label{section:intro}

Very recently, ATLAS and CMS collaborations have announced their first hosts of new results based on $3.2 \fb^{-1}$ and $2.6\fb^{-1}$ integrated luminosity at LHC Run II $\sqrt{s}=13$ TeV~\cite{ATLAS13_diphoton, CMS13_diphoton}. 
Among various channels in searching for new physics, there is an intriguing existence of diphoton excess around 750 GeV, with a local significance of $3.6$ $\sigma$ and $2.6$ $\sigma$ respectively in ATLAS and CMS. 
With more data accumulating, whether this is due to a statistical fluctuation or some manifestation of new physics would be revealed soon. 
Nevertheless, as theorists, we should always be aware if this could be the first light that changes our current understanding of microscopic physics.

The first appearing of this anomalous diphoton resonance at LHC Run II would unambiguously tell us some information. 
First, due to the Landau-Yang theorem~\cite{Yang:1950rg}, this resonance can only be spin zero or two instead of one. 
Second, the resonance decay into diphoton process can only be through the higher dimensional operators~\cite{HiggsWid}. 
Therefore, an unsuppressed total decay width would require an unconventional large production rate and one might need to try hard to hide its main decay channel into the SM backgrounds. 
Third, according to the 8 TeV LHC Run I results, CMS search~\cite{Khachatryan:2015qba} sets a 95\% CL observed upper limit of $\sigma(pp \rightarrow \Phi) {\rm Br}(\Phi \rightarrow \gamma \gamma) < {1.5}$ fb, and ATLAS search~\cite{Aad:2015mna} also imposes a similar constraints on RS gravitons. 
In order to accommodate both LHC Run I and Run II results, a larger enhancement on the diphoton signal from 8 TeV to 13 TeV is needed and the gluon initial state is preferred. 
Collecting all the above hints, we consider a singlet scalar $\Phi$ with only SM higher dimensional couplings to gluon and neutral gauge bosons as perhaps the most optimal solution.

While the process $gg \rightarrow \Phi \rightarrow \gamma \gamma$ looks simple, it does have a very rich and deep physics behind it. 
If the $\Phi$ is pseudo-scalar, or even a pseudo-Nambu-Goldstone boson (pNGB), then the only existence of higher dimensional couplings to gluon and neutral gauge bosons is a natural consequence of $\Phi$ parity violation due to anomaly~\cite{singletcoup}. 
The anomaly induced process at the IR, which affects the $\Phi$ production and decay, is proportional to the number of colour $N_n$ in the underlying confining strong dynamics. 
Therefore if this excess continues to exist in the future, by accurately measuring the diphoton resonance rate and the relative rate among different SM diboson decay channels, we could learn $N_n$ and the hypercharge of the confining vector fermions. 
This provides us another example of learning the ultraviolet physics at the infrared just like a rediscovery of colour $N_c= 3$ in QCD through $\pi_0 \rightarrow \gamma \gamma$. 
If one of the confining vector fermions $\psi$ is a SM singlet, the baryon which is made of $N_n$ copies of $\psi$ could be a composite dark matter candidate if $\psi$ is the lightest confining vector fermions. 
Therefore, we use this particular choice of charge assignment as the benchmark of our model in the phenomenology discussion.

The layout of this paper is as follows. 
In Sec.~\ref{section:EFT}, we parametrize our setup in terms of effective theory and give a numerical fit to the 13 TeV diphoton excess. 
In Sec.~\ref{sec:model}, we build up the simplest hidden QCD model with the required singlet couplings through anomaly based on $\gSU(4)$ flavor symmetry and discuss the related phenomenology. 
The benchmark model with a composite dark matter (DM) candidate from the hidden baryon is highlighted.
We make conclusions in Sec.~\ref{section:conclusion}. 
We provide an Appendix~\ref{section:WZW} to derive the effective Wess-Zumino-Witten (WZW) terms of the simplest confining model.


\section{The effective theories}
\label{section:EFT}

We first consider the general dimension-five couplings between a singlet (pseudo)scalar and the gauge fields,
\beqs\label{eq:Operators}
\beqn
&-( \frac{\alpha_s  C_g^S}{16\,\pi F }G_{\mu\nu}^a  G^{\mu\nu\,,a} + \frac{\alpha C_\gamma^S }{2\, \pi F}  F_{\mu\nu} F^{\mu\nu} )  \Phi_S\,,  \\
& - (\frac{\alpha_s C_g^P }{16\,\pi F} G_{\mu\nu}^a \widetilde G^{\mu\nu\,,a} + \frac{\alpha C_\gamma^P}{2\, \pi F}  F_{\mu\nu} \widetilde F^{\mu\nu}) \Phi_P \,,
\label{eq:eft}
\eeqn
\eeqs
respectively. 
Here, $G_{\mu\nu}^a$ and $F_{\mu\nu}$ are $\gSU(3)_c$ and $\gU(1)_{\rm em}$ field strength tensors. $F$ is expect to be the energy scale of the underlying new physics model. 
The decay widths of (pseudo)scalar $\Phi_S( \Phi_P)$ to gluon and photon pairs could be obtained from Eq.~\eqref{eq:Operators},
\beqs\label{eq:decayWid_EFT}
\beqn
\Gamma(\Phi_{S/P}\rightarrow \gamma\gamma)&=&\frac{ M_{\Phi_{S/P}}^3\alpha^2 }{16\pi^3 F^2}(C_\gamma^{S/P})^2 \,,\\
\Gamma(\Phi_{S/P}\rightarrow gg)&=&\frac{M_{\Phi_{S/P}}^3\,\alpha_s^2}{128\pi^3 F^2}(C_g^{S/P} )^2\,.
\eeqn
\eeqs

There can be different origins for obtaining the operators in Eq.~\eqref{eq:Operators}. 
The first example is analogous to the Higgs effective couplings to gluons and photons. 
One may consider a set of vector-like colour-charged and/or electric-charged fermions $F_k$ coupling with the singlet scalars such that $ \sum_k \bar F_k ( \lambda_k \Phi_S + i\gamma_5 \hat \lambda_k \Phi_P) F_k\,$.
Integrating out the heavy fermions $F_k$, the Wilson coefficients in the effective Lagrangian Eq.~\eqref{eq:eft} are calculated to be, 
\begin{eqnarray}\label{eq:vl}
C_\gamma^S&=& \sum_k \frac{\lambda_k N_c(k)Q_k^2}{3} \,,\quad
C_\gamma^P=  \sum_k \frac{\hat\lambda_k N_c(k)Q_k^2}{2}\,,\nonumber\\
C_g^S&=&\sum_k \frac{4   \lambda_k}{3} \,,\quad
C_g^P=\sum_k {2  \hat \lambda_k} \,,
\end{eqnarray}
with $N_c(k)$ and $Q_k$ being the $\gSU(3)_c$ colour degrees of freedom and charges carried by $F_k$ and we assume degenerate masses of $M_k=F$ for simplicity. 
In the following, we will denote the couplings and charges of vector-like fermion collectively as $\lambda_F=\{ \lambda_k\,,\hat \lambda_k \}$ and $Q_F=\{ Q_k \}$. 
In addition, the effective coupling for the $CP$-odd operator can be induced by the anomaly of the chiral symmetry breaking. 
This is nothing but the effective WZW term~\cite{Wess:1971yu, Witten:1983tw}. 
In the next section, a specific model construction will be given.

Due to the couplings, we expect the following relations~\footnote{Hereafter, the superscript $S,P$ of $C_{\gamma,g}$ would be dropped for simplicity, and will be identified when needed.}
\beqs
\beqn\label{eq:SP_width}
&& \Gamma(\Phi_{S/P}\rightarrow gg)\propto (C_g/F)^2\,,\\
&& \Gamma (\Phi_{S/P}\rightarrow \gamma\gamma) \propto (C_\gamma/F)^2\,.
\eeqn
\eeqs
The cross section of the production process of $gg\to \Phi_{S/P}$ should only depend on $(C_g/F)^2$, hence the following semi-analytic formula for the signal process could be expected, $\sigma[gg\to \Phi_{S/P} \to \gamma\gamma ]\sim\left(C_g/F\right)^2 (C_\gamma^2 \, f_2)/(C_\gamma^2 \, f_2 + C_g^2 \, f_1)\,$. 
Here $(f_1\,, f_2)$ are constants and one expect $f_1/f_2\propto (\alpha_s / \alpha)^2\sim (10^2- 10^3)$ (notice that the charge $Q_k$ is absorbed into $C_\gamma$. 
We find $\sigma[gg \to \Phi_{S/P}  ]\simeq (6.2 \, \fb)\times C_g^2 (\frac{1\,\TeV}{F})^2$.

\begin{figure}[htb]
\centering
\includegraphics[width=6cm,height=6cm]{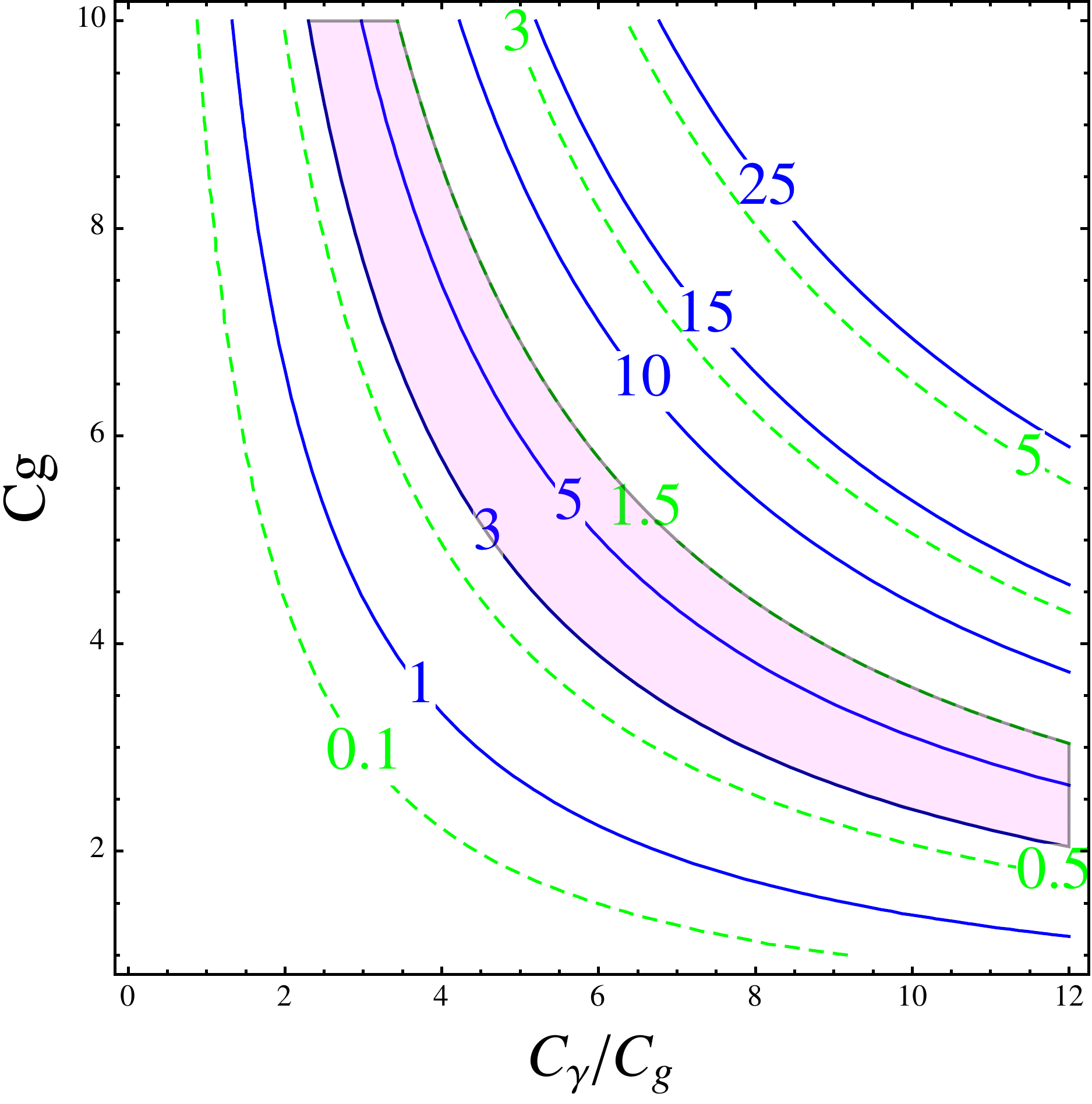}
\caption{\label{fig:diphoton_EFT} The cross sections of $\sigma[pp\to \Phi\to \gamma\gamma]$ (unit: fb) on the $(C_\gamma/C_g\,,C_g)$ plane at 8 TeV (in green dashed) and 14 TeV (in blue solid) with $F = 1$ TeV. The pink region represents the region that can explain the LHC 13 TeV results while remains unconstrained from LHC Run-I bound.}
\end{figure}

We perform the numerical analysis of the diphoton excess by using the implementation of the dimension-five operators ~Eq.~\eqref{eq:Operators} in FeynRules~\cite{Alloul:2013naa} and generate events with MadGraph~\cite{Alwall:2011uj}, interfaced with Pythia~\cite{Sjostrand:2006za} and Delphes~\cite{deFavereau:2013fsa} for the parton shower, hadronization and the fast detector simulations. 
The analysis is conducted based on the CMS cuts in Ref.~\cite{CMS13_diphoton}. 
The diphoton events are reconstructed by selecting photons such that $p_T(\gamma )\ge 75\,\GeV$, $|\eta(\gamma)|\leq 2.5$ and $|\eta (\gamma)|$ not within $(1.4442\,, 1.566)$. 
At least one photon should be in the barrel region, i.e., $|\eta|\leq 1.4442$. The diphoton invariant mass should be $m_{\gamma\gamma} \geq 230\,\GeV$. 
For events with one photon in the end-cap region such that $|\eta(\gamma)|\geq 1.566$, we require $m_{\gamma\gamma}\geq 320\,\GeV$. 
Furthermore, only the events with the diphoton invariant mass in the range of $m_{\gamma\gamma}\in (650\,\GeV\,, 800\,\GeV)$ are selected.
In order to account for the diphoton excess at LHC, the cross section of $\sigma(pp \rightarrow \Phi \rightarrow \gamma \gamma)$ at the LHC 13 TeV should be around $3-13$ fb, while the Run I constrains tell us that the cross section at the LHC 8 TeV should be less than $1.5$ fb.
Fig.~\ref{fig:diphoton_EFT} shows the contour plot on the ($C_\gamma/C_g,C_g$) plane for the $\sigma(pp \rightarrow \Phi \rightarrow \gamma \gamma)$  at the 8 TeV and 13 TeV LHC, from which we can infer that  small $C_g \lesssim 3$ requires a large ratio $C_\gamma/C_g \gtrsim 8$. 
Notice that in the case of degenerate coupling and charge  for the vector-like fermions, $C_\gamma / C_g = N_c Q_F^2/4$, therefore from the plot, we can see that our theory even remains to have small coupling $\lambda_F \lesssim 1$ for $Q_F \gtrsim 3$ when we fix $M_F = 1\,\TeV$ ($C_g^S = 4 \lambda_F N_f / 3$ and $C_g^P = 2 \lambda_F N_f$ through Eq.~\eqref{eq:vl})

\section{The model with new strong dynamics}
\label{sec:model}

Next, we turn to a specific model setup, by assuming a new QCD-like strong sector with the gauge symmetry of $\gSU(N_n)$. 
The new strong dynamics possesses the properties of confinement and the asymptotic freedom. 
We denote the pion decay constant of new strong dynamics as $f_\Pi$, and the dynamical scale as $\Lambda_n $.
They are related as $\Lambda_n \simeq 4\pi f_\Pi/\sqrt{N_n}$ by the large $N$ scaling relation.
Unlike the technicolour theories~\cite{Weinberg:1975gm,Susskind:1978ms,Hill:2002ap}, the strong sector is not necessarily related to the electroweak symmetry breaking. 
As a result, we are free to consider the case where only gauge bosons of unbroken gauge symmetries, namely, gluons and photons, can talk directly to the new sector. 
We assume a set of vector-like fermions $(\psi^1\,, \psi^2)$ under the fundamental representation of the new gauge group $\gSU(N_n)$. 
We are especially interested in the case that only the $\gSU(3)_c$ and $\gU(1)_Y$ fields have the anomaly in order to account for the diphoton excess. 
This can be realized by assuming that the fundamental fermions $(\psi^1\,, \psi^2)$  in the new strong sector belong to the singlets of SM $\gSU(2)_L$ gauge group. 
In order to embed the $\gSU(3)_c$ group and have the colour-singlet pNGBs, we consider the minimal case with $N_f = 4$ and gauge the subgoup $\gSU(3)_c \times \gU(1)_Y$ of $\gSU(4)$. 
The quantum numbers of $(\psi^1\,, \psi^2)$ are summarized in Table.~\ref{tab:field}. 
More general discussions of the global symmetry breaking patterns in different representations under the new gauge symmetries can be found in Ref.~\cite{Belyaev:2015hgo}.
The LHC phenomenology of the models with vectorlike confinement were studied in Refs.~\cite{Kilic:2009mi, Kilic:2010et}.

\begin{table}[ht]
\begin{center}
\begin{tabular}{|c|c|c|c|c|}
\hline
      & $\gSU(N_n)$ & $\gSU(3)_c$ & $\gSU(2)_L$  & $\gU(1)_Y$    \\
\hline
  $\psi^1_{L,R}$   & $N_n$ & $3$ & $1$  & {$Y_1$}    \\
\hline
  $\psi^2_{L,R}$   & $N_n$ & $1$ & $1$  & {$Y_2$}  \\
\hline
\end{tabular}
\end{center}
\caption{The field content of the minimal model. $\forall Y_{1,2}\in \mathbb{Q}$.}\label{tab:field}
\end{table}

All gauge anomalies are cancelled since the fermions $(\psi^1\,, \psi^2)$ are vector-like under both $\gSU(N_n)$ and the SM gauge groups. 
The requirement of the asymptotical freedom of the new gauge theory $\gSU(N_n)$ bounds the number of flavor to be $N_f \leq 11 N_n/2$, which is easily satisfy for $N_f=4$ and $N_n\geq 3$ in our model~\footnote{Since the fundamental representation of $\gSU(2)$ is pseudo-real, the corresponding global chiral symmetry breaking will become $\gSU(8)\to \gSp(8)$.  Hence, the meson spectra in the new strong sector will be different from the current context. }.
In addition, the asymptotical freedom of QCD should be retained, which bounds the color degrees of freedom in the new strong sector as $N_n\leq 10$.

In the limit of vanishing SM gauge couplings, the strong sector possesses the global chiral symmetry of $\gSU(4)_L \times \gSU(4)_R \times \gU(1)_A\times \gU(1)_{B_n}$, where $\gU(1)_{B_n}$ denotes the baryon number symmetry in the new strong sector.
The axial symmetry of $\gU(1)_A$ is broken by the instanton effects and will be neglected henceforth. 
The confinement of the theory at a new scale $\Lambda_n$ will induce spontaneous chiral symmetry breaking of $\gSU(4)_L \times \gSU(4)_R \rightarrow \gSU(4)_V$, which results in $15$ pNGBs $\Pi^A$. They can be decomposed into representations of SM gauge goup $\gSU(3)_c\times \gSU(2)_L\times \gU(1)_Y$ as follows
\begin{equation}
\begin{split}
15\rightarrow (8\,,1)_0  +  {(3\,,1)_{Y_1 - Y_2 }} + { (\bar{3}\,,1)_{Y_2 - Y_1} }+ (1\,,1)_0\,.
\end{split}
\end{equation}
Below, we label the colour octet, triplet and singlet pNGBs as $\Phi_8$, $\Phi_3$, and $\Phi_1$, respectively. The masses of pNGBs arise from the gauge-invariant mass terms of $- m_Q (\bar \psi^1 \psi^1 + \bar \psi^2 \psi^2 )$ for vector-like fermions. Such mass terms provide the dominant mass source and the gauging of SM group only splits the colour octets $\Phi_8$ and singlets $\Phi_1$. The meson masses are given by the Gell-Mann$-$Oakes$-$Renner relation~\cite{GellMann:1968rz}
\begin{equation}\label{eq:meson_masses}
\begin{split}
M_{\Phi_8}^2 &\sim 2m_Q\Lambda_n^3/f_\Pi^2 + \left(\frac{3g_s^2}{16\pi^2} \right)\Lambda_n^2\,,\\
M_{\Phi_3}^2 &\sim 2m_Q\Lambda_n^3/f_\Pi^2 + \left(\frac{g_s^2}{24\pi^2} + \frac{(Y_1-Y_2)^2\,g^{\prime 2}}{16\pi^2} \right)\Lambda_n^2\,,\\
M_{\Phi_1}^2 &\sim 2m_Q\Lambda_n^3/f_\Pi^2 \,.
\end{split}
\end{equation}
The singlet remains the lightest as a result of the Witten theorem~\cite{Witten:1983ut} and is a good candidate for the diphoton excess.  
Below, we focus on a benchmark model with $f_\Pi\simeq 2.5\,\TeV$ with $N_n=10$ for a composite DM candidate in the spectrum.
The corresponding vector-like fermion mass is found to be $m_Q\simeq 1\,\GeV$ to accommodate a $750\,\GeV$ singlet in the new strong dynamical sector.
Note, the benchmark model given here is made by assuming the unitarity bound to the DM mass is saturated. 
More generic choices of the pion decay constant of $f_\Pi$ and the gauge symmetries can be expected.
Correspondingly, the vector-like fermion mass range spanning from $\mO(1)\,\GeV$ to $\mO(10)\,\GeV$ is also natural.

The pion-number violating interactions arise from the effective WZW term
\begin{equation}\label{eq:WZWterm}
\begin{split}
\mathcal{L}_{\rm WZW} = - \frac{N_n \,g_B g_C}{8\, \pi^2}\frac{\Pi_A}{f_{\Pi}} F^{\mu\nu\,,B}\widetilde F_{\mu\nu}^C \text{Tr}[T^A T^B T^C]\,,
\end{split}
\end{equation}
where $g_{B,C}$ are the gauge couplings associated with the SM gauge field strength tensors $F^{B,C}_{\mu\nu}$, and the dual field strength tensor is defined as $\widetilde F_{\mu\nu}^A\equiv \hf \epsilon_{\mu\nu\rho\sigma }F^{\rho\sigma\,,A}$.
Here, the trace is performed over the SM indices.

In principle, we can freely choose the hypercharges of $\gSU(3)_c$ triplet and singlet. 
The singlet couplings to the hypercharge fields lead to the possible diphoton signals, given that $Y_1\neq Y_2$. 
Note that topological interactions between the colour-triplet and the SM gauge bosons are forbidden by the $\gSU(3)_c$ symmetry. 
An interesting and special case is $Y_2=0$ where $\psi^2$ is a SM singlet. 
In this case, the baryonic composites $(\psi^2)^{N_n}\equiv \epsilon^{a_1\,...a_{N_n}} \psi_{a_1}^2\,... \psi_{a_{N_n}}^2$ could be a dark matter candidate. 
Therefore, we will use $Y_2=0$ as our benchmark model in the later discussions.

\begin{figure}[htb]
\centering
\includegraphics[width=5.5cm,height=4.5cm]{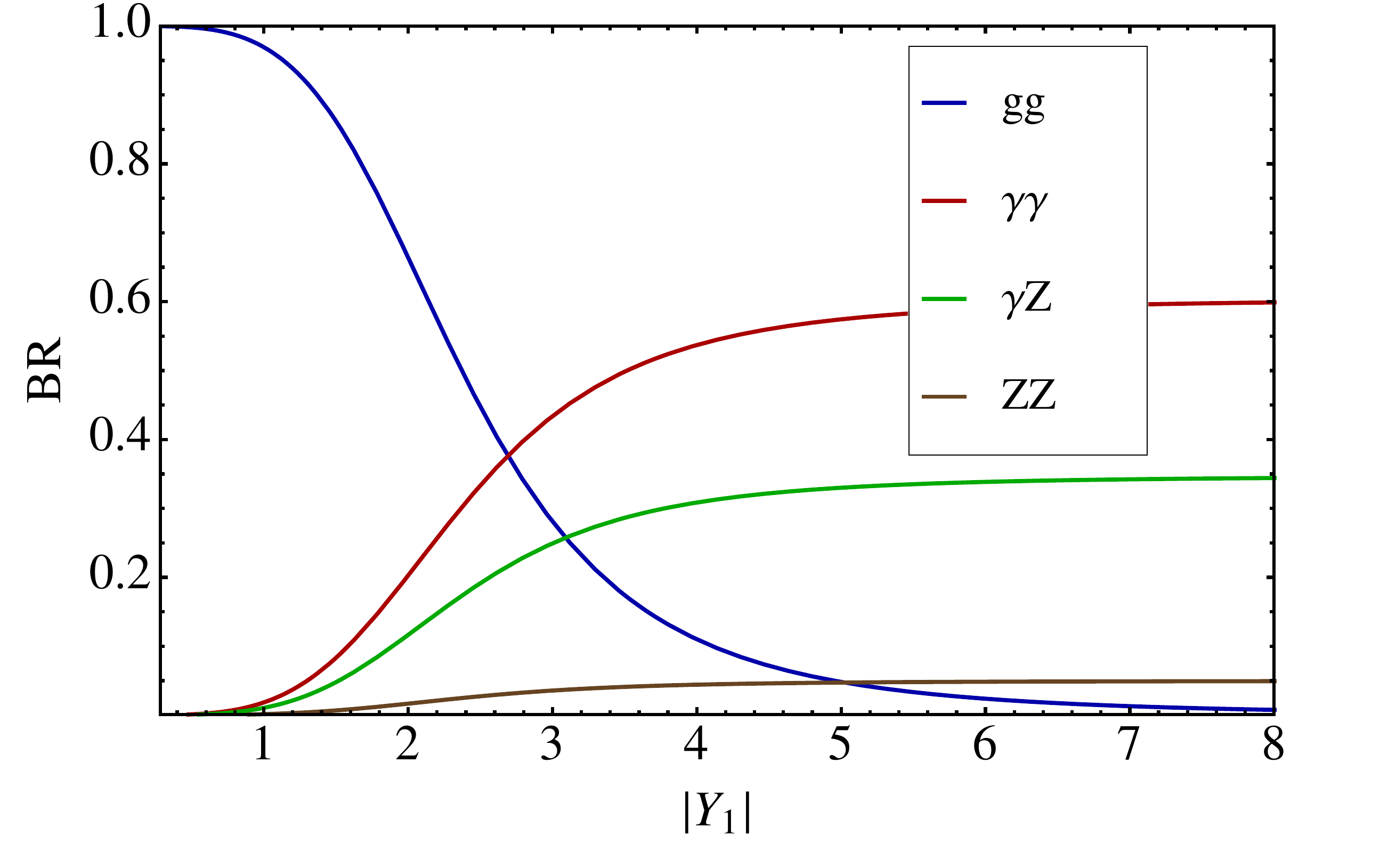}
\caption{\label{fig:phi1_decay} The decay branching ratios into various final states for the singlet $\Phi_1$. 
}
\end{figure}

According to the effective WZW terms in Eq.~\eqref{eq:WZWterm} for the singlet $\Phi_1$, we expect its couplings to the $(gg\,, \gamma\gamma\,, \gamma Z\,, ZZ)$ final states. Likewise, the colour octet $\Phi_8$ can couple to $(gg\,,g \gamma\,, gZ)$. 
The explicit expressions for the WZW effective terms are given in Eqs.~\eqref{eq:phi1_WZW} and \eqref{eq:phi8_WZW}, with the corresponding partial decay widths of  $\Phi_1$ and $ \Phi_8$ given in Eqs.~\eqref{eq:singlet_width} and \eqref{eq:octet_width}.
In Fig.\ref{fig:phi1_decay}, we display the decay branching fractions of the singlet $\Phi_1$ with the varying inputs of $|Y_1|$ by assuming that $Y_2=0$.
The increasing $|Y_1|$ apparently leads to the enhancement of the decay modes of $(\gamma\gamma\,, \gamma Z\,, ZZ)$.

If the future LHC experiments discover both singlet $\Phi_1$ and octet $\Phi_8$, one may measure the signal rates to $\gamma\gamma$ and dijets for $\Phi_1$, and the signal rates of $\gamma$ plus jets and dijets for $\Phi_8$. 
Obviously, these signal rates can be used to determine the hypercharges in the minimal model. 
From Eqs.~\eqref{eq:singlet_width} and \eqref{eq:octet_width}, one has
\begin{equation}
\begin{split}
Y_1^2 &=\frac{5}{96} \frac{\alpha_s}{\alpha} \frac{\Gamma[\Phi_8 \to g \gamma] }{\Gamma[\Phi_8 \to gg ]}\,,\\
Y_2^2 &= Y_1^2 \pm \frac{\alpha_s}{\alpha} \Big( \frac{2}{9}\cdot \frac{\Gamma[\Phi_1 \to \gamma \gamma] }{\Gamma[\Phi_1 \to g g] }  \Big)^{1/2}\,.
\end{split}
\end{equation}
Note from the meson mass spectrum in Eq.~\eqref{eq:meson_masses}, the octets are typically $\mO(1)\,\TeV$ heavier than the singlets.

\begin{figure}[ht]
\centering
\includegraphics[width=5.5cm,height=4.5cm]{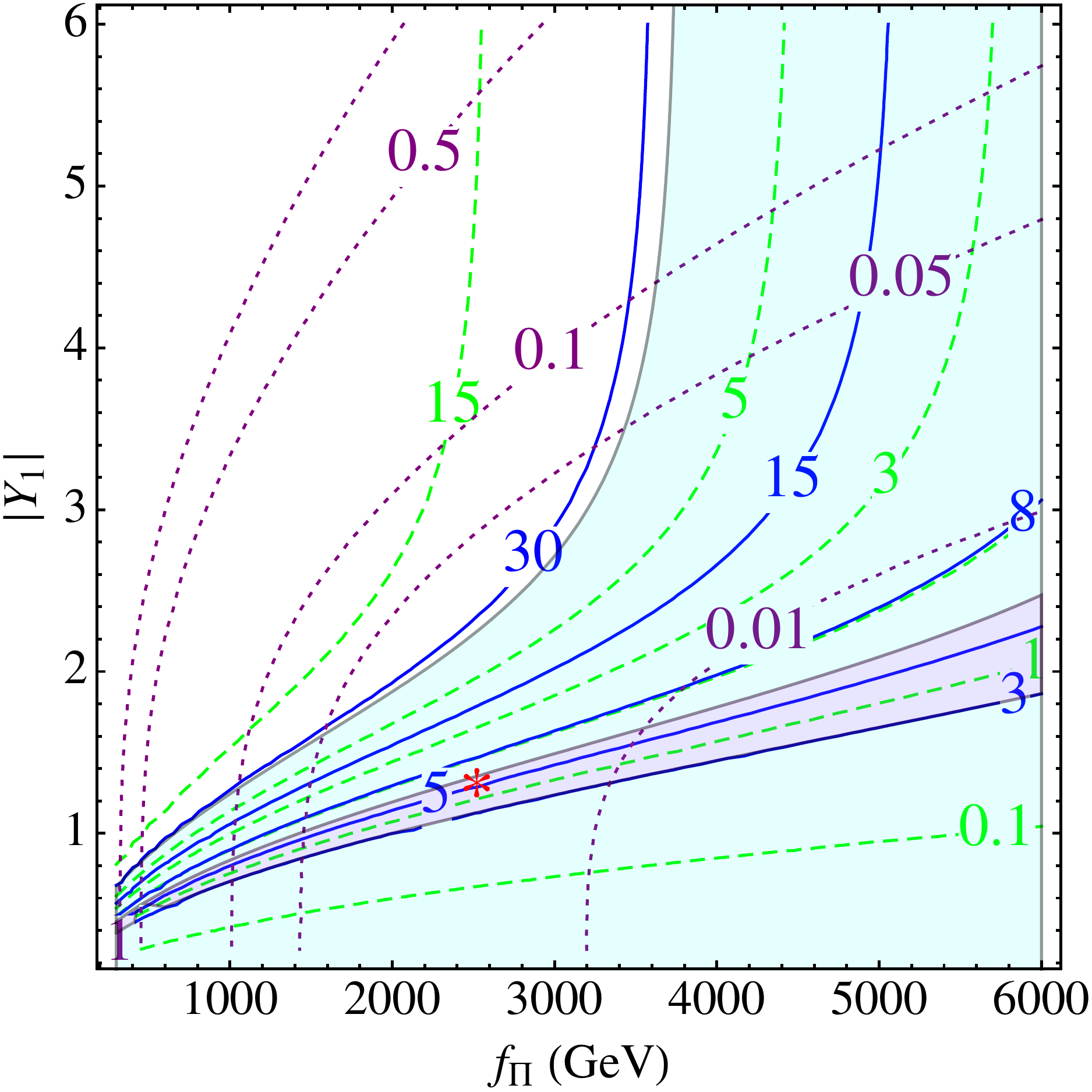}
\caption{\label{fig:singlet_diphoton} The diphoton cross section of $\sigma[p p \to \Phi_1\to \gamma\gamma]$ (unit: fb) on the $(f_\Pi\,, |Y_1|)$ plane for $Y_2=0$, with $K=1.6$~\cite{Spira:1995rr} for the NLO QCD correction. The blue and green curves correspond to the 13 TeV and 8 TeV predictions, respectively. The total decay width of $\Phi_1$ in unit of GeV is shown by purple contours. Cyan region are allowed region of $Z\gamma$ final states from CMS Run I, and magenta region is favored by 13 TeV diphoton excess and allowed by 8 TeV CMS diphoton limits. The colour number in the hidden strong sector is chosen as $N_n=10$. A benchmark point marked by star correspond to $f_\Pi\simeq 2.5\,\TeV, |Y_1|=4/3$, and the diphoton signal cross section of $5.4\,\fb$. }
\end{figure}

The total production cross sections can be obtained by mapping the EFT parameter $C_g$ into the minimal model, which is $C_g= \frac{2N_n}{\sqrt{6}}$ from Eqs.~\eqref{eq:WZWterm} and \eqref{eq:phi1_WZW}. 
{\it It is crucial to note the production of the colour singlet $\Phi_1$ is proportional to the number of colours in the new confining strong dynamics.} 
Based on our numerical simulations by following the approaches described in the previous section and the decay branching ratios displayed in Fig.~\ref{fig:phi1_decay}, we plot the signal predictions of $\sigma[p p \to \Phi_1\to \gamma\gamma]$ on the $(f_\Pi\,, |Y_1|)$ plane in Fig.~\ref{fig:singlet_diphoton} by fixing $Y_2=0$ for $\gSU(10)$ hidden gauge symmetry. 
The parameter region with large $|Y_1|$ and small $f_\Pi$ inputs have been excluded by the ATLAS searches for $Z\gamma$~\cite{Aad:2014fha}.
Furthermore, the total decay width of $\Phi_1$ is found to be hardly larger than $0.1$ GeV for parameter regions favored by the LHC 13 TeV and 8 TeV data sets. 
Therefore, the limits on cross section of the spin-0 resonance to diphoton in the narrow width hypothesis with $\Gamma_X=0.1$ GeV~\cite{Khachatryan:2015qba} is applicable to some extent.

In addition to the diphoton signal predictions we explored above, the minimal model also predicts several other experimental signatures. 
We note above that the colour singlet $\Phi_1$ decays also to $gg$. 
Correspondingly, one would envision the future dijet searches around the mass resonance of $\sim 750\,\GeV$. 
Expressed in terms of the effective couplings defined in Eq.~\eqref{eq:Operators}, the ratio between the diphoton and gluon pair signals are determined by $ (C_\gamma/ C_g)^2 = \frac{9}{16 }(Y_1^2 - Y_2^2)^2$. 
Therefore, the future observation and measurements of the dijet signals are not only useful for justifying the model, but also crucial for determining the hypercharge differences for the underlying model.

The other experimental constraints related to the minimal model are the searches for the vector-like quarks $\psi^1$ and the colour triplet $\Phi_3$ in the spectrum. They can be pair produced and hadronize with quarks and gluons to form the $R$-hadrons. 
The $R$-hadron searches at the LHC thus place mass limits on the colour triplet, and the colour triplet mass is bound above around 845 GeV by LHC Run-I results~\cite{Chatrchyan:2013oca,ATLAS:2014fka}. 
The LHC Run-II at 13 TeV would set the bound more stringent~\cite{Barnard:2015rba}. 
The analysis in Ref.~\cite{Liu:2015bma} set an exclusion to charged stable particles to $\sim 900\,\GeV$ for sufficiently long decay lengths of $c\tau\geq 10$ m.

The colour triplet $\Phi_3$ has hypercharge $Y_1-Y_2$, so that it can decay to SM lepton-quark pair through higher-dimensional operators for specific choices of $Y_1 - Y_2$. Examples include
\begin{equation}\label{eq:phi3decay}
\begin{split}
Y_1 - Y_2 =  \frac{2}{3}&: \frac{ 1}{ \Lambda^{2} } (\bar \psi^2 \gamma^\mu\gamma_5 \psi^1) (\bar{d}_{R} \gamma_\mu e_{R})\,,\\
Y_1 - Y_2 = \frac{5}{3} &: \frac{ 1}{ \Lambda^{2} } (\bar \psi^2 \gamma^\mu\gamma_5 \psi^1) (\bar{u}_{R} \gamma_\mu e_{R})\,.
\end{split}
\end{equation}
For the benchmark in Fig.~\ref{fig:singlet_diphoton}, one may look for the pair-produced lepton-quark signals of $jj\ell \ell$ to search for this resonance. 
The current mass exclusion of $\Phi_3$ is $\sim 1\,\TeV$ from the ATLAS 8 TeV searches~\cite{Aad:2015caa}.
For sufficiently small $\Lambda$, the metastable $\Phi_3$ is expected to decay before the era of the big bang nucleosynthesis (BBN)~\cite{Nakai:2015ptz,Harigaya:2016pnu,Kawasaki:2004qu} through Eq.~\eqref{eq:phi3decay}.
This is different from the case in Ref.~\cite{Appelquist:2015yfa} where small symmetry-breaking Yukawa couplings between the new fermions and the Higgs doublet are expected to avoid the tension with BBN induced by the metastable charged meson. 
We left the careful study of BBN to future work.

There are also baryonic composites in the $\gSU(N_n)$ confining  gauge theory since $\pi_3( \gSU(4)^2/ \gSU(4) ) = \mathbb{Z}$. 
Such baryonic composites usually saturate the cut off scale $4 \pi f_\Pi$ and thus are heavy. 
They are also the topological objects, which in general get more suppressed production rate besides of the heavy mass kinematical suppression. 
Therefore, we only expect very tiny production rate at the LHC. 
In the case of $m_{\psi^1} \geqslant m_{\psi^2}$ and $Y_2 $= 0~\footnote{If $Y_2$ is nonzero, we can use higher dimensional operators to decay $(\psi^2)^{N_n}$.}, the $(\psi^2)^{N_n}$ baryonic composites would be a composite DM candidate. 
The thermally averaged annihilation rate of the dark baryons could be estimated using partial wave unitarity~\cite{Appelquist:2015yfa,Griest:1989wd,Blum:2014dca}, and the dark baryon mass could be bounded from above as $m_B\lesssim 100\,\TeV$~\footnote{It is also likely that an asymmetric relic density may extend the DM mass range even below few $\TeV$s~\cite{Appelquist:2015yfa}.}.
When this bound is saturated, we find $f_\Pi\simeq 2.5\,\TeV$ with $N_n=10$ as our benchmark model marked by star in Fig.~\ref{fig:singlet_diphoton}. 
The estimation is made by employing the large $N$ scaling for the composite baryon mass of $m_B\sim N_n \Lambda_n$~\cite{Witten:1979kh, Antipin:2015xia}.
In the early universe, once it is thermally produced, the correct abundance of baryonic dark matter could be obtained with a relatively strong coupling. One may ask whether this baryonic dark matter is metastable through instanton effects~\cite{D'Hoker:1983kr}. However, the enormous suppression factor which is proportional to exp$(-8 \pi^2/g_s^2)$ will make its life time much longer than the age of our universe. And the consequence of topological dark matter on cosmic ray signals and the decay of the DM through higher dimension operators is highly related with the choice of $N_n$~\cite{Huo:2015nwa, Murayama:2009nj}, which is beyond the scope of this work.
For more generic case without considering a hypothetical composite DM candidate with $Y_2=0$ as in the minimal model, the 750 GeV diphoton signals can be accommodated by varying $f_\Pi$ from several hundred GeV to $\mO(1)\,\TeV$ with $3\leq N_n \leq 10$.




\section{Conclusion}
\label{section:conclusion}

We have studied the possibility that a singlet scalar ($CP$-even or $CP$-odd) to account for the recent diphoton excess observed by ATLAS and CMS which has attracted a lot of interests~\cite{Pilaftsis:2015ycr, Angelescu:2015uiz, Franceschini:2015kwy, Buttazzo:2015txu, Nakai:2015ptz, Harigaya:2015ezk, Mambrini:2015wyu, Knapen:2015dap, Backovic:2015fnp, DiChiara:2015vdm,Higaki:2015jag,McDermott:2015sck,Ellis:2015oso,Low:2015qep,Bellazzini:2015nxw,Gupta:2015zzs,Petersson:2015mkr,Molinaro:2015cwg,Ahmed:2015uqt, Bai:2015nbs,Aloni:2015mxa,Falkowski:2015swt,Csaki:2015vek,Chakrabortty:2015hff,Curtin:2015jcv,Fichet:2015vvy,Chao:2015ttq,Demidov:2015zqn,No:2015bsn,Becirevic:2015fmu,Martinez:2015kmn,Agrawal:2015dbf,Cox:2015ckc,Kobakhidze:2015ldh,Matsuzaki:2015che,Cao:2015pto}. We focused on the gluon-gluon initiated process and studied it in an effective field theory approach, where the corresponding higher dimension operators can be generated through the heavy vector-like fermions. We then consider a natural example that the singlet scalar is a pseudo-Goldstone boson from chiral symmetry breaking of a new strong sector and the interactions with SM gauge bosons (the gluon, photon and $Z$) are purely topological and arising from anomaly. We consider the minimal flavor symmetry group $\gSU(4) \times \gSU(4)$ with 15 pNGBs and find the lightest pNGB is colour-singlet, which is a good candidate for the diphoton excess. Our model also predicts the colour-octet, colour-triplet scalars and composite baryons. For the colour-triplet, there is no topological interaction arising from anomaly due to the $\gSU(3)_c$ symmetry, therefore it could be meta-stable due to higher dimensional operators and generates $R$-hadron like signal at the LHC. Other anomaly decays of colour-octet scalars are also discussed. The lightest neutral baryon could be a viable composite dark matter. If the diphoton excess is confirmed in the near future, the discovery of these resonances and the precise production and decay will provide us a strong test on our scenario.


\noindent{\bfseries Acknowledgments.} We thank Yang Bai, Ran Huo, Zhen Liu, and Lian-tao Wang for very helpful discussions. This work is supported by National Science Foundation of China (under grant No. 11335007, 11575176), and the Fundamental Research Funds for the Central Universities (under grant No. WK2030040069 and No. 0903005203404).


\smallskip
\smallskip
\noindent{\it Note added.} Before this paper is submitted, Ref.~\cite{Nakai:2015ptz} appeared which also discussed a singlet pNGB production and decay through anomaly. Nevertheless, our model considers a more general charge assignment and focus on the case where the second confining fermion $\psi^2$ is a SM singlet, where $N_n$ copies of $\psi^2$ could be a composite dark matter candidate. We also consider the possibility that colour triplet scalar is meta-stable, which results a $R$-hadron like signal at the LHC Run II (its decay pattern through the higher dimensional operator is also different). Moreover, we calculate the predicted hypercharge of the two confining fermions and the confining colour number $N_n$ in terms of pNGBs different diboson decay rate.

\appendix
\section{ The $\gSU(4)$ generators and WZW effective terms}
\label{section:WZW}

We use the generalized Gell-Mann matrices as our generators
\begin{equation}
\begin{split}
T^a &= \frac12\left(\begin{array}{cc}
\lambda^a & 0\\
0 & 0
\end{array}
\right)\,, \qquad
T^{15}= \frac{1}{2\sqrt{6}}\left(\begin{array}{cccc}
1& 0&0&0\\
0 & 1 &0 &0\\
0 & 0 &1 &0\\
0& 0 &0 &-3\\
\end{array}
\right)\\
\end{split}
\end{equation}
where $\lambda^a$ for $a = 1, \cdots, 8$ are the $\gSU(3)$ Gell-Mann matrices with $\Tr(\lambda^a \lambda^b)= 2 \delta^{ab}$, and $T^{15}$ is the third Cartan generator. 
$T^i$ with $i=9\,,...14$ are not used in our evaluation and their expressions are neglected. 
Collectively, we write down the pNGBs as
\beqn
\Pi_A T^A&=& \Phi_8^a T^a + \Phi_3^i \hat T^i + (\Phi_3^i \hat T^i  )^\dag + \Phi_1 T^{15}\,.\nonumber\\
\eeqn
Our hypercharge generator is defined by
\begin{equation}
Y\equiv \text{diag}(Y_1\,, Y_1\,, Y_1\,, Y_2)\,.
\end{equation}

With the conventions listed above, it is straightforward to write down the WZW term between the pNGB $\Pi_A$ and the SM gauge fields according to Eq.~\eqref{eq:WZWterm}.
For the singlet $\Phi_1$, they read
\begin{equation}\label{eq:phi1_WZW}
\begin{split}
\mL_{\Phi_1}&=  - \frac{N_n g_s^2}{ 8 \pi^2}\frac{\Phi_1 }{f_{\Pi}} G_{\mu\nu}^a \widetilde G^{b\,,\mu\nu}\Tr [T^{15} T^a T^b ] \\
& - \frac{N_n g^{\prime2}}{8 \pi^2}\frac{\Phi_1  }{f_{\Pi}}  B_{\mu\nu}\widetilde B^{\mu\nu} \Tr [ T^{15} Y^2 ] \\
&= - \frac{ N_n \alpha_s }{8\sqrt{6} \pi f_\Pi } \Phi_1 G_{\mu\nu}^a \widetilde G^{a\,,\mu\nu} \\
&-\sqrt{6} (Y_1^2 - Y_2^2 ) \frac{N_n \alpha}{8\pi f_\Pi} \Phi_1 ( \widetilde A_{\mu\nu} A^{\mu\nu} \\
& - 2 t_W \widetilde Z_{\mu\nu } A^{\mu\nu} + t_W^2 \widetilde Z_{\mu\nu} Z^{\mu\nu}) \,,
\end{split}
\end{equation}
where $t_W\equiv \sin\theta_W/\cos\theta_W $.
For octets $\Phi_8$, they read
\begin{equation}\label{eq:phi8_WZW}
\begin{split}
\mL_{\Phi_8}&= - \frac{N_n g_s^2 }{8\pi^2} \frac{\Phi_8^a }{f_\Pi} G_{\mu\nu}^b \widetilde{G}^{c\,,\mu\nu} \Tr[T^a T^b T^c] \\
&- \frac{N_n g^{\prime } g_s }{8\pi^2 } \frac{\Phi_8^a }{f_\Pi} G_{\mu\nu}^b \tilde B^{\mu\nu} \Tr[T^a T^b Y] \\
&= - \frac{N_n \alpha_s }{8\pi f_\Pi}\, d^{abc} \, \Phi_8^a G_{\mu\nu}^b \widetilde G_{\mu\nu}^c \\
&- \frac{ Y_1 N_n \sqrt{\alpha\alpha_s} }{4 \, \pi  f_\Pi} \Phi_8^a G_{\mu\nu}^a ( - t_W \widetilde Z^{\mu\nu} + \widetilde A^{\mu\nu} ) \,,
\end{split}
\end{equation}
where the symmetric tensor $d^{abc}$ is given by
\beqn\label{eq:dsymbol}
\{ T^a\,, T^b \}&=& \frac{1}{3} \delta^{ab} + d^{abc} T^c\,.
\eeqn

The partial decay widths for the singlet $\Phi_1$ can be obtained from the WZW term between the pNGB $\Pi_A$ and gauge fields from Eq.~\eqref{eq:phi1_WZW},
\begin{equation}\label{eq:singlet_width}
\begin{split}
\Gamma[\Phi_1 \to gg]&= \frac{N_n^2 \alpha_s^2 }{ 192\pi^3 f_\Pi^2} M_{\Phi_1}^3  \,,\\
\Gamma[\Phi_1 \to \gamma\gamma]&= \frac{3 N_n^2 \alpha^2   }{ 128 \pi^3 f_\Pi^2} (Y_1^2 - Y_2^2 )^2 M_{\Phi_1}^3 \,,\\
\Gamma[\Phi_1\to \gamma Z]&=  \frac{3 N_n^2 \alpha^2 t_W^2 }{64\, \pi^3 f_\Pi^2} (Y_1^2 - Y_2^2 )^2 M_{\Phi_1}^3\,\\
&\times \Big( 1- \frac{m_Z^2}{M_{\Phi_1}^2}  \Big)^3 \,,\\
\Gamma[\Phi_1 \to ZZ ]&= \frac{3 N_n^2 \alpha^2 t_W^4 }{128\,\pi^3 f_\Pi^2} (Y_1^2 - Y_2^2 )^2 M_{\Phi_1}^3\, \\
&\times \Big(  1- \frac{4 m_Z^2}{M_{\Phi_1}^2} \Big)^{3/2} \,,
\end{split}
\end{equation}
and also the partial widths of the octet $\Phi_8$ from Eq.~\eqref{eq:phi8_WZW},
\begin{equation}\label{eq:octet_width}
\begin{split}
\Gamma [\Phi_8 \to gg] &= \frac{5}{4} \frac{M_{\Phi_8}^3}{ M_{\Phi_1}^3} \Gamma[\Phi_1 \to gg] \,,\\
\Gamma[\Phi_8 \to g\gamma] &= 24 \frac{\alpha\,Y_1^2}{\alpha_s} \frac{M_{\Phi_8}^3}{M_{\Phi_1}^3} \Gamma[\Phi_1 \to gg] ,\\
\Gamma[\Phi_8 \to gZ] &=24 \frac{\alpha t_W^2\, Y_1^2}{\alpha_s} \frac{M_{\Phi_8}^3}{M_{\Phi_1}^3} \\
&\times \Big( 1-\frac{m_Z^2}{ M_{\Phi_8}^2} \Big)^3 \Gamma[\Phi_1 \to gg]\,.\\
\end{split}
\end{equation}
%



\begin{thebibliography}{999}
\bibitem{HiggsWid}   
The apparent width in the diphoton invariant mass distribution can be just detector effects like in the 125 GeV Higgs discovery.
\bibitem{singletcoup}
 In other models where a singlet scalar couples to the vector fermions (charged under $\gSU(3)_c$ and $\gU(1)_Y$), the coupling between $\Phi$ and SM gauge bosons are loop induced. To get the required excess rate, the coupling between $\Phi$ and vector fermion can be weak if the vector fermion has a a moderately large $Y_f$. This can exist in many perturbative models with singlet, like NMSSM, or some dark matter models, etc.
\end{thebibliography}
\end{document}